\documentclass[showpacs,preprintnumbers,amsmath,amssymb]{revtex4}

\usepackage{graphicx}% Include figure files
\usepackage{dcolumn}% Align table columns on decimal point
\usepackage{bm}% bold math
\newcommand {\apgt} {\ {\raise-.5ex\hbox{$\buildrel>\over\sim$}}\ }
\newcommand {\aplt} {\ {\raise-.5ex\hbox{$\buildrel<\over\sim$}}\ }

\begin{document}
 
\title{Conceptually new mechanism for trapping neutral, polar 
         particles. }

\author{R. Bl\"umel}  
\affiliation{Department of Physics, Wesleyan University, 
Middletown, Connecticut 06459-0155}
  
\date{\today}

\begin{abstract}
It is shown that 
a superposition of static and rapidly oscillating electric {\it monopole} (source) 
fields is 
capable of trapping particles with a permanent electric dipole moment. 
Thus, the new trapping 
mechanism differs fundamentally from saddle-point traps that 
use static and oscillating higher-multipole fields. 
An analytical stability analysis together with detailed 
molecular dynamics trajectory calculations prove 
that the trap is stable. Thin rods of barium titanate (BaTiO$_3$)
provide an illustrative example for the working principle of the new trap.  
The effects of gravity are 
considered. The existence of a bifurcation regime is predicted. 
A particular strength of the new trap is that it also works for 
zero orbital angular momentum with respect to the 
field-generating electrodes. 
  
\end{abstract}

  % Electrets 
                       
%\pacs{05.45.-a, %nonlinear dynamics and nonlinear dynamical systems
%      32.80.Pj} %optical cooling of atoms; trapping


\maketitle

%%%%%%%%%%%%%%%%%%%%%%%%%%%%%%%%%%%%%%%%%%%%%%%%%%%
\section{Introduction}
Levitation of microscopic and macroscopic objects is 
of general interest in physics: It is a key technology 
for basic and applied sciences. For instance, research areas 
ranging from high-resolution spectroscopy 
\cite{HRSpec1,HRSpec2,HRSpec3}, 
frequency standards \cite{FrequStand}, 
neutral anti-matter production \cite{NAM1,NAM2}, 
Bose-Einstein condensation \cite{JILABEC,MITBEC}, 
and quantum 
computing \cite{QComp} to high-speed trains \cite{MagLev}, all rely on the 
levitation and three-dimensional confinement of objects 
by electromagnetic fields. The key point is that 
electromagnetic confinement provides ``walls of pure energy'' 
that avoid contamination, friction, and, in the case of 
antimatter, annihilation, 
by avoiding contact with material walls. 
 
Traps used in science and technology may broadly 
be divided into charged-particle traps and 
neutral-particle traps. Historically, charged-particle traps 
were developed first, 
starting with the development of the 
Kingdon trap \cite{Kingdon1} in the 1920s, the 
Penning trap \cite{Penning} 
in the 1930s 
and the Paul trap \cite{Paul1}  
in the 1950s. In addition to their use in 
high-resolution spectroscopy 
\cite{HRSpec1,HRSpec2,HRSpec3}, 
frequency standards \cite{FrequStand}, 
and quantum 
computing \cite{QComp}, 
these traps were used, e.g., 
as a model for planetary dynamics 
\cite{orr}, 
for studying collisions and excited-state lifetimes of 
ions \cite{Church}, and the production 
and investigation of Coulomb crystals 
\cite{MPQ-PRL,NIST-PRL,RB-Nature}. 
Recently, neutral-particle 
traps were developed and demonstrated as well. 
Based on the type of field employed, we 
may distinguish several different types of 
neutral-particle traps, for instance, electric 
\cite{WHW,Seka,SekaSchm,BethNat,VBM,BVSM,MWtrap}, 
magnetic \cite{Lovelace,Berge}, or light-field 
traps \cite{BIO,ASH}. 
 
The purpose of this paper is to introduce a conceptually new 
trap for the stable confinement of neutral, polar particles. 
The trapping mechanism of the new trap is fundamentally different from 
the trapping mechanisms employed in existing neutral-particle 
traps: (i) The trap does not make use of energy shifts in 
selected quantum states and 
(ii) in contrast 
to existing traps \cite{VBM,BVSM}, which are based on a dynamic 
saddle-point potential akin to the strong-focusing Paul-trap 
principle \cite{Paul1}, 
the new trap works on the basis of 
a superposition of static and oscillating {\it monopole} fields. 
  
The paper is organized in the following way.
In Sec.~\ref{Equations of motion} we derive the coupled 
radial and angular equations of 
motion of point dipoles trapped 
in a combination of ac and dc electric monopole (source) fields. 
In Sec.~\ref{Linear stability analysis} we analyze the stability 
of the coupled system of equations and derive conditions 
for the stable confinement 
of point dipoles. We conclude that stable trapping 
is possible as long as stable solutions 
of the radial equation exist. Therefore,  
in Sec.~\ref{Properties of the radial equation}, 
we investigate in detail the stability properties 
of the radial equation. We find that 
subject to certain conditions on the control parameter of the 
radial equation, the radial equation is stable, 
which guarantees the stability 
of the combined radial and angular system of equations 
derived in Sec.~\ref{Equations of motion}, and thus results in 
stable confinement of point dipoles according to the 
new trapping scheme. 
In Sec.~\ref{NumIn}, strengthening our 
analytical results, we show numerically that the trap is stable. 
In Sec.~\ref{Beating gravity} we derive the conditions 
under which stable confinement in the presence of 
gravity is achieved. 
In Sec.~\ref{Design algorithm} we state a step-by-step 
algorithm according to which trap parameters may be 
designed. 
The trapping 
of thin, cylindrical 
rods of barium titanate is discussed in 
Sec.~\ref{Ferroelectric rods}. 
In Sec.~\ref{Discussion} we discuss our results. In 
Sec.~\ref{Summary and conclusions}  we summarize our results 
and conclude the paper.
 
%==============================================================================
 
\section{Equations of motion} 
\label{Equations of motion}
In this section we derive the equations of motion of a point 
dipole ${\bm p}$ of mass $M$ and moment of inertia $I$ in an electric field 
\begin{equation}
{\bm E}(r,t)\ =\ \frac{\nu V(t)}{r^{\alpha}}\, \hat{\bm r}, 
\label{EQMO1}
\end{equation}
where $\nu>0$ and $\alpha>0$ are constants, $r$ 
is the distance of the dipole from the source of the field, 
$\hat{\bm r}$ is the unit vector in ${\bm r}$ direction, 
and $V(t)$ is the voltage applied 
to the electrodes that generate the field 
${\bm E}$.   
Since the 
interaction energy of a point dipole ${\bm p}$ with an electric 
field ${\bm E}$ is 
\begin{equation}
W\ =\ -{\bm p}\cdot {\bm E}\ =\ -p E \cos(\theta),
\label{EQMO2}
\end{equation}
where $\theta$ is the angle between ${\bm p}$ and 
${\bm E}$, 
the Lagrangian function ${\cal L}$ of the point 
dipole in the case of zero orbital angular momentum is 
\begin{equation}
{\cal L}\ =\ \frac{1}{2} M \dot r^2\, +\, \frac{1}{2} I [\dot\theta^2\, +\, 
\sin^2(\theta)\dot\varphi^2]\, +\, \frac{p\nu V(t)}{r^{\alpha}}\, \cos(\theta),
\label{EQMO3}
\end{equation}
where $\varphi$ is the azimuthal angle of ${\bm p}$. 
Since $\partial {\cal L}/\partial\varphi = 0$, 
the azimuthal angle is a cyclic variable \cite{LL}, and the 
conjugate canonical momentum, 
\begin{equation}
L_{\varphi}\ =\ \frac{\partial {\cal L}}{\partial \dot\varphi}\ =\ I\sin^2(\theta) 
\dot\varphi, 
\label{EQMO4}
\end{equation}
is a constant of the motion. In the following we specialize to the case 
$L_{\varphi}=0$. While some neutral-particle traps require 
non-zero orbital angular momentum to work properly 
(an example is the trap described in \cite{Seka,SekaSchm}), 
a particular strength of the trap introduced in this paper 
is that it works for zero orbital angular momentum. Therefore, 
playing to the strength of the new trap, we treat the case 
in which 
the dipole is not rotating around 
the field source, i.e. the orbital angular momentum of the point dipole 
around the field-generating electrodes is zero. It has been checked 
by means of extensive numerical simulations that the trap also 
works for non-zero orbital angular momentum. However, since this 
case does not add substantial additional insight into the fundamental 
working principle of the new trap, the case of non-zero orbital 
angular momentum is not treated in this paper and we will 
continue to focus on the case of zero orbital angular momentum 
as reflected in the Lagrangian function (\ref{EQMO3}). 
On the basis 
of the Lagrangian function (\ref{EQMO3}) 
the radial ($r$) and angular ($\theta$) equations of motion are 
obtained from the Lagrangian equations 
\cite{LL} 
\begin{equation}
\frac{d}{dt}\, \frac{\partial {\cal L}}{\partial \dot r}\, -\, 
\frac{\partial{\cal L}}{\partial r}\ =\ 0, \ \ \ 
\frac{d}{dt}\, \frac{\partial {\cal L}}{\partial \dot \theta}\, -\, 
\frac{\partial{\cal L}}{\partial \theta}\ =\ 0. 
\label{EQMO5}
\end{equation}
From (\ref{EQMO3}) with (\ref{EQMO5}) we obtain 
\begin{align}
&\ddot r\, +\, \frac{p\alpha\nu V(t)\cos(\theta)}{M r^{\alpha + 1}}\ =\ 0, 
\label{EQMO6a}     \\
&\ddot\theta\, +\, \frac{p\nu V(t)\sin(\theta)}{I r^{\alpha}}\ =\ 0. 
\label{EQMO6b}
\end{align}
We now specialize the voltage $V(t)$ to 
\begin{equation}
V(t)\ =\ V_{\rm dc}\, -\, V_{\rm ac}\cos(\Omega t),
\label{EQMO7}
\end{equation}
where $V_{\rm dc}>0$ is the dc part of the voltage, 
$V_{\rm ac}>0$ is the ac part of the voltage, and 
$\Omega$ is the angular frequency of the ac part of 
the voltage. We also define the dimensionless 
control parameters 
\begin{equation}
\eta\ =\ \frac{V_{\rm ac}}{2 V_{\rm dc}} 
\label{EQMO8}
\end{equation}
and 
\begin{equation}
 \beta\ =\ \frac{M l_0^2}{\alpha I}, 
\label{EQMO14}
\end{equation}
the unit of time 
\begin{equation}
t_0\ =\ \frac{2}{\Omega} 
\label{EQMO9}
\end{equation}
and the unit of length
\begin{equation}
l_0\ =\ \left( \frac{4\alpha p \nu V_{\rm dc}}{M\Omega^2}\right)^{\frac{1}{\alpha + 2}}, 
\label{EQMO10}
\end{equation}
such that 
\begin{equation}
\rho\ =\ \frac{r}{l_0}
\label{EQMO11}
\end{equation}
and 
\begin{equation}
 \tau\ =\ \frac{t}{t_0}
\label{EQMO12}
\end{equation}
are the dimensionless radius and time, respectively. 
In terms of $\rho$ and $\tau$, with (\ref{EQMO8}) -- 
(\ref{EQMO12}), the dimensionless 
equivalents of (\ref{EQMO6a}) and (\ref{EQMO6b}) are 
\begin{align}
&\ddot \rho\, +\, [1-2\eta\cos(2\tau)]\frac{\cos(\theta)}{\rho^{\alpha+1}}\ =\ 0, 
\label{EQMO13a}    \\
&\ddot\theta\, +\, \beta [1-2\eta\cos(2\tau)]\frac{\sin(\theta)}{\rho^{\alpha}}\ =\ 0,  
\label{EQMO13b}
\end{align}
where the dots now indicate differentiation with 
respect to $\tau$. 

%==============================================================================
 
\section{Linear stability analysis}   
\label{Linear stability analysis} 
It is not obvious that the equations of motion 
(\ref{EQMO13a}) and (\ref{EQMO13b}) have stable solutions. 
Indeed, we will see in this and following sections, that 
both $\eta$ and $\beta$ 
need to satisfy certain conditions in order for 
(\ref{EQMO13a}) and (\ref{EQMO13b}) to exhibit stable 
solutions. To find the condition for $\beta$, we 
perform a linear stability analysis. We emphasize that 
expanding the equations of motion to linear order is 
not an uncontrolled approximation. 
In the absence of pathologies -- as is the case here -- it is an exact method 
for assessing the stability properties of the equilibrium solution 
of the equations of motion \cite{Ott}. 
   
To first order in $\theta$, we may replace 
$\cos(\theta)$ by 1 to obtain 
\begin{equation}
\ddot \rho\, +\, [1-2\eta\cos(2\tau)]\frac{1}{\rho^{\alpha+1}}\ =\ 0.
\label{LSA0}
\end{equation}
Thus, to linear order, the $\rho$ equation 
decouples from the $\theta$ equation. This fact substantially 
simplifies the stability analysis. However, the 
$\theta$ equation  
remains coupled to the $\rho$ equation 
since even 
in linear order the 
$\theta$ equation contains $\rho$. 
To lowest order the Fourier expansion of 
the exact solution $\rho(\tau)$ of 
(\ref{LSA0}) 
is given by 
\begin{equation}
\rho(\tau)\ =\ \rho_0\, -\, \rho_1\cos(2\tau),
\label{LSA1}
\end{equation}
where $\rho_0$ and $\rho_1$ are constants. 
Inserting (\ref{LSA1}) into (\ref{LSA0}), 
replacing $\cos^2(2\tau)$ terms by their average 
value $1/2$, and neglecting 
$(\alpha+1)\rho_1/\rho_0$ with respect to $2\eta$, 
the $\rho$ equation (\ref{LSA0}) 
is fulfilled if 
\begin{align}
\rho_0\ &=\ \left( \frac{\alpha + 1}{2}\right)^{\frac{1}{\alpha+2}}\, 
\eta^{\frac{2}{\alpha+2}}, 
\label{LSA2a}   \\
\rho_1\ &=\ \left[ 2^{\frac{1}{\alpha+2}}\, 
(\alpha+1)^{\frac{\alpha+1}{\alpha+2}}\, 
\eta^{\frac{\alpha}{\alpha+2}} \right]^{-1}.
\label{LSA2b}
\end{align}
Inserting (\ref{LSA1}) into (\ref{EQMO13b}), keeping only terms 
linear in $\rho_1$, and replacing once more all $\cos^2(2\tau)$ 
terms 
by $1/2$, we obtain 
\begin{equation}
\ddot\theta\ +\ \frac{\beta}{\rho_0^{\alpha}}\, 
\left\{ \left(\frac{1}{\alpha+1}\right)\, -\, 
2\eta\left[ 
1-\frac{\alpha}{2\eta^2(\alpha+1)}\right]\, \cos(2\tau)\right\}\, 
\sin(\theta)\ =\ 0. 
\label{LSA3}
\end{equation}
Furthermore, 
since we are focusing on small oscillations, we may 
safely replace $\sin(\theta)$ by $\theta$. In addition, 
as we will see below, $\eta$ is usually quite large 
($\eta \apgt 10$) 
so that 
we may neglect the term $\alpha/[2\eta^2(\alpha+1)]$ 
with respect to 1. Together, this yields 
\begin{equation}
\ddot\theta\ +\ \frac{\beta}{\rho_0^{\alpha}}\, 
\left[ \left(\frac{1}{\alpha+1}\right)\, -\, 
2\eta \cos(2\tau)\right]\, 
\theta\ =\ 0. 
\label{LSA4}
\end{equation}
This is a Mathieu equation \cite{AS}, whose canonical 
form is \cite{AS} 
\begin{equation}
\ddot x\ +\ 
\left[ a\, -\, 2q\cos(2\tau)\right]x\, =\, 0. 
\label{LSA5}
\end{equation}
Comparing (\ref{LSA5}) with (\ref{LSA4}) we see that 
\begin{align}
a\ &=\ \frac{\beta}{(\alpha+1)\rho_0^{\alpha}}, 
\label{LSA6a}   \\
q\ &=\ \frac{\beta\eta}{\rho_0^{\alpha}}. 
\label{LSA6b}
\end{align}
It is well known \cite{AS} that stable solutions of (\ref{LSA5}) exist only 
in certain two-dimensional regions of the $(q,a)$ parameter plane \cite{AS}. 
Because of $\rho_0>0$ and 
the definitions of $\alpha$ and $\beta$ in (\ref{EQMO1}) and 
(\ref{EQMO14}), respectively, we have $a>0$. 
In this case, and up to second order in $q$, the first, and 
most important, stability region of the Mathieu equation is bounded 
by \cite{AS} 
\begin{equation}
0\ <\ a\ <\ 1\, -\, q\, -\, \frac{q^2}{8}.
\label{LSA7}
\end{equation}
With (\ref{LSA6a}) and (\ref{LSA6b}) we have 
\begin{equation}
a\ =\ \frac{q}{(\alpha+1)\eta} . 
\label{LSA8}
\end{equation}
Using this in (\ref{LSA7}), we obtain the 
following quadratic inequality for $q$: 
\begin{equation}
q^2\, +\, 8q\left[ 1+\frac{1}{(\alpha+1)\eta}\right]\, -\, 8\ <\ 0.
\label{LSA9}
\end{equation}
Keeping only terms up to first order in 
$1/[(\alpha+1)\eta]$, the solution is 
\begin{equation}
q\ <\ q_s(\eta),
\label{LSA10}
\end{equation}
where 
\begin{align}
q_s(\eta)\ &=\ (-4+2\sqrt{6})\, -\, \frac{4}{(\alpha+1)\eta}\left[ 
1-\frac{\sqrt{6}}{3}\right] \\
&=\ 0.9\, -\, \frac{0.734}{(\alpha+1)\eta}.
\label{LSA11} 
\end{align}
With (\ref{LSA2a}) and (\ref{LSA6b}) this yields the 
stability criterion 
\begin{equation}
\beta\ <\ q_s(\eta)\frac{\rho_0^{\alpha}}{\eta}\ =\ 
q_s(\eta)\, \left(\frac{\alpha+1}{2}\right)^{\frac{\alpha}{\alpha+2}}\, 
\eta^{\frac{\alpha-2}{\alpha+2}}.
\label{LSA12}
\end{equation}
In summary, a range of $\beta$ values exists that leads to stable 
$\theta$ oscillations. This result 
is significant. It means that as long as 
stable solutions of (\ref{LSA0}) exist, 
stable solutions of the system (\ref{EQMO13a}) and (\ref{EQMO13b}) 
can be constructed, which, in turn, implies stable confinement 
of dipoles. Therefore, before 
proceeding further, we need to investigate the 
stability properties of the radial equation (\ref{LSA0}). 
   
%==============================================================================
 
\section{Properties of the radial equation} 
\label{Properties of the radial equation}  
A pseudo-potential analysis \cite{Dehmelt} is the most natural way to 
conduct a 
comprehensive stability analysis of the radial equation (\ref{LSA0}). 
It is a powerful method that may be applied generally for the analysis 
of the motion of particles in rapidly oscillating force fields \cite{LL}. 
According to this method, we start with (\ref{LSA1}), but instead of 
holding $\rho_0$ fixed, we allow $\rho(\tau)$ to execute slow 
oscillations around $\rho_0$. Therefore, we now write 
\begin{equation}
\rho(\tau)\ =\ R(\tau)\, -\, \rho_1\cos(2\tau),
\label{PREQ1}
\end{equation}
where, as in (\ref{LSA1}), $\rho_1$ is a constant. 
Equation (\ref{PREQ1}) represents formally what actually happens 
physically: The exact motion $\rho(\tau)$ is a superposition of a 
slow, large-amplitude motion $R(\tau)$ and a fast, small-amplitude 
motion proportional to $\cos(2\tau)$ \cite{LL}. The slow, large-amplitude 
motion is known as the 
macro-motion; the fast, small-amplitude motion is the micro-motion. 
Apparently, because it determines the cycle-averaged trajectory 
of a trapped particle, it is the behavior of $R(\tau)$, 
which determines the 
stability properties of the trap. 
 
Inserting (\ref{PREQ1}) into (\ref{LSA0}), keeping only 
terms up to first order in $\rho_1$, and, as we did before, 
replacing $\cos^2(2\tau)$ by its average value $1/2$, we obtain 
\begin{align}
&\left[ \ddot R\, +\, \frac{1}{R^{\alpha+1}}\, -\, 
\frac{\eta(\alpha+1)\rho_1}{R^{\alpha+2}} \right] \, +\, 
\nonumber    \\ 
&\left\{ 4\rho_1\, -\, \frac{1}{R^{\alpha+1}} \left[ 2\eta-(\alpha+1)
\frac{\rho_1}{R} \right] \right\}\, \cos(2\tau)\ =\ 0. 
\label{PREQ2}
\end{align}
This equation is fulfilled if, separately, 
\begin{align}
 \ddot R\, +\, \frac{1}{R^{\alpha+1}}\, -\, 
\frac{\eta(\alpha+1)\rho_1}{R^{\alpha+2}}\, &=\, 0, 
\label{PREQ3a}   \\ 
4\rho_1\, -\, \frac{1}{R^{\alpha+1}} \left[ 2\eta-(\alpha+1)
\frac{\rho_1}{R} \right] \ &=\ 0. 
\label{PREQ3b}
\end{align}
Neglecting the term $(\alpha+1)\rho_1/R$ in (\ref{PREQ3b}) with 
respect to $2\eta$, equation (\ref{PREQ3b}) yields 
\begin{equation}
\rho_1\ =\ \frac{\eta}{2 R^{\alpha+1}}.
\label{PREQ4}
\end{equation}
Inserting this result into (\ref{PREQ3a}) yields 
\begin{equation}
\ddot R\ =\ -\frac{1}{R^{\alpha+1}}\, +\, \frac{(\alpha+1)\eta^2}{2R^{2\alpha+3}}\ =\ 
-\frac{\partial V_{\rm eff}(R)} {\partial R},
\label{PREQ5}
\end{equation}
where 
\begin{equation}
V_{\rm eff}(R)\ =\ -\frac{1}{\alpha R^{\alpha}}\, +\, \frac{\eta^2}{4 R^{2\alpha+2}} 
\label{PREQ6}
\end{equation}
is known as the effective potential \cite{LL}, the pseudo-potential \cite{Dehmelt}, or 
the ponderomotive potential in atomic \cite{Friedrich} and plasma \cite{Morales} 
physics. Apparently, the motion 
of the point dipole, on average, behaves as if the dipole were under the influence 
of a force that is the gradient of the pseudo-potential $V_{\rm eff}(R)$. This is 
a powerful result that allows us to determine the stability of the motion of 
the point dipole according to physical reasoning: If $V_{\rm eff}(R)$ has a 
potential minimum, the point dipole will execute stable, oscillatory motion 
inside of the potential minimum; the point dipole is trapped. 
If $V_{\rm eff}$ does not have a potential 
minimum, the point dipole will escape to infinity, and no stable trapping is 
observed. A potential minimum occurs at points $R_0$ for which 
$V'_{\rm eff}(R_0)=0$ and $V''_{\rm eff}(R_0)>0$. Indeed, 
$V'_{\rm eff}(R_0)=0$ has a single solution for 
\begin{equation}
R_0\ =\ \rho_0,
\label{PREQ7}
\end{equation}
where $\rho_0$ is defined in (\ref{LSA2a}).
Apparently, linear stability analysis and 
pseudo-potential analysis are consistent in that they 
predict the same equilibrium radius $\rho_0$. 
Moreover, we have 
\begin{equation}
V''_{\rm eff}(R_0)\ =\ \frac{2(\alpha+2)}{(\alpha+1)\eta^2}\ >\ 0, 
\label{PREQ8}
\end{equation}
i.e., $R_0=\rho_0$, in fact, corresponds to a 
pseudo-potential minimum for all 
control parameters $\eta$. At this point we have 
to remember that in our 
pseudo-potential analysis we neglected higher-order terms 
in $\rho_1/R$. Thus, we expect the pseudo-potential 
analysis to be trustworthy only for $\rho_1/R_0\ll 1$, which 
implies 
\begin{equation}
\eta\ \gg\ \frac{1}{\alpha+1}. 
\label{PREQ9}
\end{equation}
Physically, the condition $\rho_1/R_0\ll 1$, or, equivalently, $R_0\gg\rho_1$ 
means that the amplitude $\rho_1$ of the fast oscillations around $R_0$ 
needs to be small enough with respect to $R_0$ such that the dipole 
does not hit the field-generating source at $r=0$. This, however, would indeed 
occur for $R_0<\rho_1$ and lead to the ejection of the 
point dipole from the trap. Indeed, there is ``trouble'' for small $\eta$. 
It is known \cite{K1} that for $\alpha=0$ 
the radial equation (\ref{LSA0}) exhibits 
a period-doubling bifurcation at 
$\eta_1\approx 3.12$ and further period-doubling bifurcations, 
following the Feigenbaum scenario \cite{Ott}, at $\eta_{j+1}<\eta_j$, 
$j=1,2,\ldots$, terminating in a transition to fully developed chaos 
at $\eta_{\infty}\approx 2.91$. 
This means that in $\eta_{j+1}<\eta<\eta_j$, $j=1,2,\ldots$, 
the lowest Fourier component 
of $\rho(\tau)$ is not $\cos(2\tau)$, as in (\ref{LSA1}), but 
$\cos(2\tau/2^j)$. This is not taken into account 
in (\ref{LSA1}). As a consequence, our pseudo-potential analysis 
is not valid for $\eta<\eta_1$. Detailed calculations 
show that (i) the 
period-doubling scenario persists for $\alpha > 0$, and (ii) that 
$\eta_1(\alpha)$ is a slowly, monotonically decreasing function 
of $\alpha$, ranging from 
$\eta_1(\alpha=0)\approx 3.12$ to $\eta_1(\alpha=10)\approx 2.41$. 
It is important to emphasize that the occurrence 
of bifurcations does not preclude stable trapping. Indeed, stable 
trapping, in principle, may be achieved in the entire range 
$[\eta_{\infty}<\eta<\eta_1]$. Thus, the existence and 
location of the predicted bifurcations may 
be explored experimentally. However, since in this region, as 
mentioned above, the point dipole is already very close to 
the field-generating source(s), it may be very difficult to trap 
dipoles experimentally in the bifurcation regime. 
 
Another source of trouble for the pseudo-potential analysis 
are resonances, which occur for $\eta>\eta_1$. 
Resonances occur whenever the frequency of 
oscillations in the pseudo-potential well $V_{\rm eff}$ is a 
rational multiple of the driving frequency of the trap. 
Luckily, only the lowest order resonances are ``dangerous''. 
They occur whenever the low-frequency oscillations 
of $R(\tau)$ in (\ref{PREQ1}) 
are 1/3 or 1/2 of the driving frequency ($=2$) 
in (\ref{LSA0}). 
Although the pseudo-potential analysis, incorrectly, 
predicts stability in these resonant cases, 
it is not entirely useless.  
Since $\eta>\eta_1$, the conditions for the computation 
of the pseudo-potential are met. Therefore, 
there is nothing wrong with computing the 
frequency of slow oscillations of the dipole in the 
pseudo-potential $V_{\rm eff}$ and, on the basis of this result, 
to predict the $\eta$ values at which the resonant instabilities 
occur. Since the frequency $\omega$ of small oscillations around 
the equilibrium radius $\rho_0$ is 
\begin{equation}
\omega\ =\ \sqrt{V''_{\rm eff}(\rho_0)}, 
\label{PREQ10}
\end{equation}
the instability at $\omega=2/3$ is predicted to occur at 
\begin{equation}
\eta_3^*\ =\ \frac{3}{2}\sqrt{\frac{2(\alpha+2)}{\alpha+1}}, 
\label{PREQ11}
\end{equation}
and the instability at $\omega=1/2$ is predicted to occur at 
\begin{equation}
\eta_4^*\ =\ 2\sqrt{\frac{2(\alpha+2)}{\alpha+1}}, 
\label{PREQ12}
\end{equation}
where
we used the expression (\ref{PREQ8}) for 
$V''_{\rm eff}(\rho_0)$. For $\alpha=1$, we predict 
$\eta_3^*=3\sqrt{3}/2\approx 2.6$ and 
$\eta_4^*=2\sqrt{3}\approx 3.5$ 
in fair agreement with the numerically obtained values 
$\eta_3^*\approx 3.16$ and $\eta_4^*\approx 3.86$, respectively 
\cite{K6}. More important than the numerical proximity of 
these special values is their behavior as a function of $\alpha$. 
Equations (\ref{PREQ11}) and (\ref{PREQ12}) show that both 
$\eta_3^*$ and $\eta_4^*$ are bounded and never exceed 
$\eta=4$. 
 
Summarizing the results of this and the 
previous section, we obtain the following general result: 
In the absence of gravity, stable trapping of point 
dipoles may be achieved for all $\eta > 5$. Even for 
$\eta<5$, ranges of $\eta$ values exist for which 
stable trapping is possible. The 
$\eta<5$ regime exhibits rich 
dynamics, but is less accessible to analytical 
analysis. 
%==============================================================================
 
\section{Numerical investigation of stability} 
\label{NumIn}  
In this section we back up the analytical calculations with 
numerical simulations. We solved 
the system of equations 
(\ref{EQMO13a}) and (\ref{EQMO13b}) numerically 
for $\eta=100$, $\beta=0.59$, and $\alpha=2$ 
for the following $5929$ initial conditions: $\rho(\tau=0)=13+j\times 0.2$, 
$j=-5,\ldots,5$, 
$\dot\rho(\tau=0)=k\times 0.01$, 
$k=-3,\ldots,3$, 
$\theta(\tau=0)=l\times 0.03$, 
$l=-5,\ldots,5$, 
$\dot\theta(\tau=0)=m\times 0.003$, 
$m=-3,\ldots,3$.
Since the solutions of the system of equations (\ref{EQMO13a}) and (\ref{EQMO13b}) 
are well behaved, no special numerical integrator needs to be chosen. 
We chose a simple 4th order Runge-Kutta integrator \cite{AS}, which yields 
completely converged, numerically exact solutions for any prescribed accuracy 
(we achieved a relative accuracy of better than $10^{-6}$ in all 
dynamical variables). 
Using the Runge-Kutta integrator, 
each of the 5929 initial conditions was propagated forward in time,  
and for each trajectory 
it was verified graphically that it stayed bounded 
over a time interval $\Delta \tau=500\pi$. 

\begin{figure}
\includegraphics[width=120mm,height=110mm]{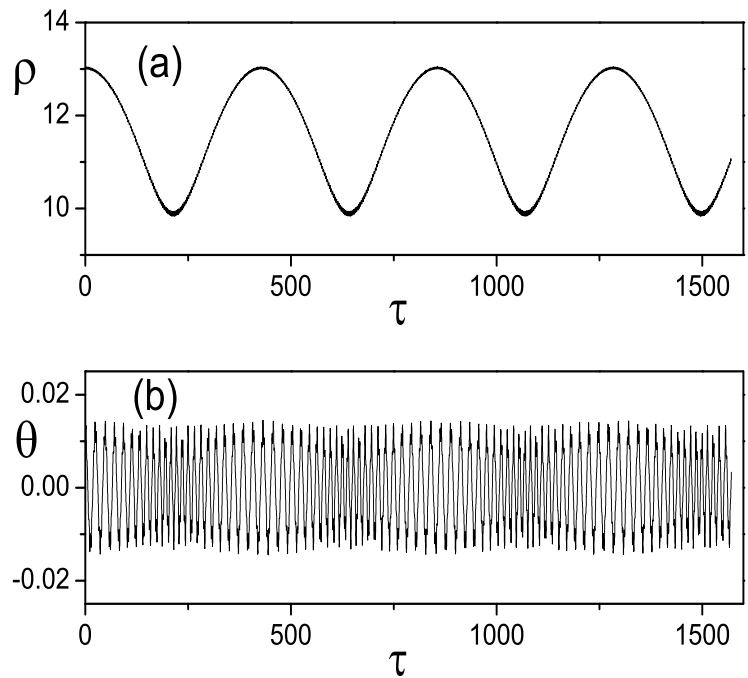}
% [scale=0.75, angle=40]
\caption{\label{fig1}  Molecular trajectory calculation of (a) $\rho(\tau)$ 
                                     and (b) $\theta(\tau)$ for 
                                     initial conditions $\rho(0)=13$, $\dot\rho(0)=0$, 
                                     $\theta(0)=0.01$, and $\dot\theta(0)=0$. The blurring 
                                     of the line in (a) results from the micro-motion, 
                                     which is not resolved on the scale of (a), and, as (a) shows, 
                                     is very small. Panel (b) shows 
                                     bounded oscillations of $\theta$ as a function of time. 
                                     Together, (a) and (b) illustrate the stability of the trap. 
                                     The calculations are numerically exact (converged) 
                                     solutions of the equations of motion 
                                     (\ref{EQMO13a}) and (\ref{EQMO13b}). 
                                     No approximations 
                                     were performed. 
   }
\end{figure}
\begin{figure}
\includegraphics[width=120mm,height=100mm]{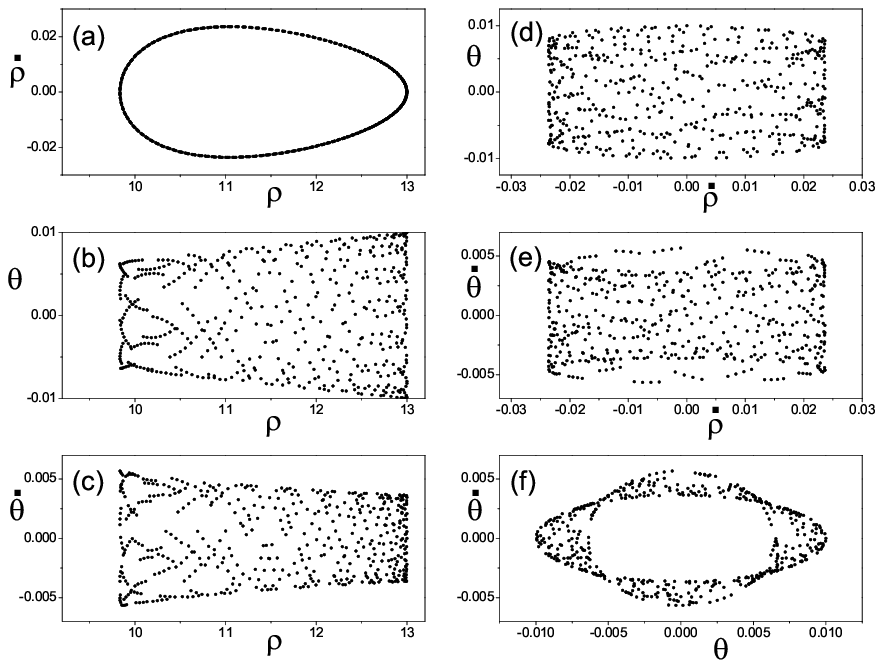}
\caption{\label{fig2} Six stroboscopic sections of the 
                                    4-dimensional phase space 
                                    of the molecular-dynamics trajectory of 
                                    Fig.~\ref{fig1} for 500 cycles of 
                                    the trap field. Each point in the figure represents 
                                    the respective values of the trajectory at the end 
                                    of each cycle. 
                                    (a) $\dot\rho$ vs. $\rho$, 
                                    (b) $\theta$ vs. $\rho$, (c) $\dot\theta$ vs. $\rho$, 
                                    (d) $\theta$ vs. $\dot\rho$, (e) $\dot\theta$ vs. $\dot\rho$, 
                                    (f) $\dot\theta$ vs. $\theta$. The six sections 
                                    show clearly that (i) the molecular trajectory is 
                                    bounded in all four phase-space dimensions, and 
                                    (ii) executes bounded 
                                    oscillations in both $\rho$ and $\theta$ directions.  
}
\end{figure}

A representative molecular trajectory 
is illustrated in Figs.~\ref{fig1} and 
\ref{fig2}. The trajectory picked is the one 
with initial conditions $\rho(0)=13$, $\dot\rho(0)=0$, 
$\theta(0)=0.01$, and $\dot\theta(0)=0$. Figure~\ref{fig1} (a) shows 
the time-evolution of $\rho(\tau)$ over the time interval 
$0\leq \tau\leq 500\pi$. During this time interval, the 
macro-motion 
performs about three and a half oscillations, ranging between 
$\rho_{\rm min}=9.8$ and $\rho_{\rm max}=13$. Thus, 
the oscillations are bounded in the $\rho$ direction. It is 
important to notice that the micro-motion is very small. 
In fact, the micro-motion amplitude is so small that it is 
not resolved on 
the scale of Fig.~\ref{fig1} (a) and manifests 
itself as a fine blur of the line. 
This is an important consistency check, since the pseudo-potential 
method is valid only if the micro-motion amplitude is 
small compared to the oscillations of the macro-motion. 
This is clearly born out in Fig.~\ref{fig1} (a). Since $\eta$ is large, 
the small amplitude of the micro-motion 
observed in Fig.~\ref{fig1} (a) is 
consistent with its analytical 
prediction (\ref{LSA2b}). 
 
Figure~\ref{fig2} illustrates the stability of the 
representative molecular trajectory in phase space. 
This is 
a more stringent test of stability than the $\rho(\tau)$ and 
$\theta(\tau)$ trajectories displayed in Fig.~\ref{fig1}, since 
the trajectory, e.g., might be stable in $\rho$, but unstable in $\dot\rho$. 
We show the molecular trajectory over the same time interval 
chosen in Fig.~\ref{fig1}, i.e. 
$\Delta\tau=500\pi$. 
Since the molecular trajectory is embedded in a 4-dimensional 
phase space ($\rho(\tau),\dot\rho(\tau),\theta(\tau),\dot\theta(\tau)$), 
we use the technique of Poincar\'e sections \cite{Ott} to display 
the trajectory via six panels of two-dimensional phase-space 
projections. We choose the projections on the six possible
two-dimensional coordinate planes 
(a) $\dot\rho$ vs. $\rho$, 
(b) $\theta$ vs. $\rho$, (c) $\dot\theta$ vs. $\rho$, 
(d) $\theta$ vs. $\dot\rho$, (e) $\dot\theta$ vs. $\dot\rho$, 
(f) $\dot\theta$ vs. $\theta$. While the micro-motion is 
essential for generating the trapping forces, it is 
the macro-motion that determines the stability 
of the trap (see Sec.~\ref{Properties of the radial equation}). 
Therefore, suppressing the 
irrelevant micro-motion and focusing on the 
relevant macro-motion, 
we strobe the motion, 
i.e., we display the trajectory at the end of each cycle 
of the trap field 
at the discrete times $\tau_j=j\times\pi$, $j=1,\ldots,500$ 
(plot symbols in Fig.~\ref{fig2}).  
Figure~\ref{fig2} (a) shows clearly that the trajectory 
executes a stable oscillation in $\rho$ direction. 
Figure~\ref{fig2} (b) shows that the motion is also bounded 
in the $\rho$-$\theta$ plane. 
Figure~\ref{fig2} (c) shows new information; we see that 
the motion is also bounded in $\dot\theta$ direction. 
Figures~\ref{fig2} (d) and (e) contain the new information 
that the motion is also bounded in $\dot\rho$. 
Figure~\ref{fig2} (f) shows that the trajectory executes 
stable oscillations in the $\theta$ direction. The six panels 
taken together imply that there is no ``escape direction'' 
in phase space, i.e. the motion is globally stable. Plots akin 
to Figs.~\ref{fig1} and \ref{fig2} were produced for each of 
the $5929$ initial conditions and overlaid one of top of 
the other. The resulting composite plots look 
qualitatively like Figs.~\ref{fig1} and \ref{fig2}, showing no 
data points outside a confined area. This way it is 
possible to summarily assess stability of all 
$5929$ molecular trajectories, without having to 
inspect them one-by-one individually. The net result is 
a numerical confirmation of stability of the trap in an entire 
four-dimensional phase-space volume. 
   
Comparing Fig.~\ref{fig1} (b) with Figs.~\ref{fig2} (b), (d), and (f), one 
might perceive a problem: The $\theta$ amplitude in Fig.~\ref{fig1} (b) is 
larger than the $\theta$ spread in Figs.~\ref{fig2} (b), (d), and (f). 
The reason for this, however, is straightforward. 
It was checked explicitly that the reduction of spread in 
Figs.~\ref{fig2} (b), (d), and (f) is due to the 
strobing of the motion at multiples of $\pi$. Strobing 
does not display the motion over a continuous time 
interval, but samples the motion at 
discrete times. This explains the small reduction in spread. 

The numerical 
check performed 
amounts to a stability survey of a four-dimensional 
phase-space volume surrounding the point 
$(\rho(\tau=0),\dot\rho(\tau=0), \theta(\tau=0), 
\dot\theta(\tau=0))$ with a resolution of 5929 initial 
conditions that all lead to stable solutions of the 
system of equations (\ref{EQMO13a}) and (\ref{EQMO13b}). 
The composite phase-space plot produced by overlaying 
all 5929 individual phase-space plots 
reveals stable 
radial oscillations in the $\rho$ interval 
$9 < \rho < 19$ and the $\theta$ interval 
$-0.3 < \theta < 0.3$. 
 
Since dipole forces are usually much smaller than 
corresponding forces on isolated electric charges, 
we anticipate that consideration and inclusion 
of gravitational forces may be important for a proper 
description of the dynamics of trapped 
dipoles according to the new trapping scheme. 
We will address the influence of gravity next. 
%==============================================================================
    
\section{Beating gravity} 
\label{Beating gravity}  
In previous sections we established that,  
in the absence of gravity, control 
parameter combinations $(\eta,\beta)$ exist that result 
in the stable, permanent confinement of point dipoles. 
Since in this paper we are focusing on electric fields of the form 
(\ref{EQMO1}), which vanish for large $r$, it is obvious 
that in certain trapping regimes 
gravity may overwhelm the confining effect of the 
electric field and result in a destabilization of 
the trap. Therefore, in this section, we study 
the effects of gravity and state the conditions that 
result in stable trapping in the presence of gravity. 

In SI units, and according to (\ref{PREQ6}), 
the pseudo-potential $V_{\rm eff}(r)$ is given by 
\begin{equation}
V_{\rm eff}(r)\ =\ \left( \frac{M l_0^2}{t_0^2}\right)\, 
\left[-\frac{1}{\alpha\rho^{\alpha}}\, +\, \frac{\eta^2}{4\rho^{2\alpha+2}} \right], 
\label{BG1}
\end{equation}
where $\rho=r/l_0$ is defined in (\ref{EQMO11}). In the presence of gravity, 
$V_{\rm eff}(r)$ is modified to $V_{\rm eff}(r)\rightarrow V_{\rm eff}(r)-Mgr$, i.e. 
\begin{equation}
V_{\rm eff}(r)\ =\ \left( \frac{M l_0^2}{t_0^2}\right)\, V_{\rm eff}(\rho), 
\label{BG2}
\end{equation}
where 
\begin{equation}
V_{\rm eff}(\rho)\ =\ 
-\frac{1}{\alpha\rho^{\alpha}}\, +\, \frac{\eta^2}{4\rho^{2\alpha+2}}\, -\, 
\gamma\rho
\label{BG3}
\end{equation}
with 
\begin{equation}
\gamma\ =\ \frac{g t_0^2}{l_0}. 
\label{BG4}
\end{equation}
For stable trapping to occur, $V_{\rm eff}(\rho)$ needs to exhibit a 
potential minimum, i.e. we require 
\begin{equation}
V'_{\rm eff}(\rho)\ =\ f(\rho)\, -\, \gamma\ =\ 0,
\label{BG5}
\end{equation}
where 
\begin{equation}
f(\rho)\ =\ \frac{1}{\rho^{\alpha+1}}\, -\, \frac{(\alpha+1)\eta^2}{2\rho^{2\alpha+3}}. 
\label{BG6}
\end{equation}
Since $\gamma>0$, equation (\ref{BG5}) has a solution only if the maximum of 
$f(\rho)$ exceeds $\gamma$. From $f'(\rho)=0$ we determine that the maximum 
of $f(\rho)$ occurs at 
\begin{equation}
\rho_{\rm max}\ =\ \left( \frac{2\alpha+3}{2}\right) ^{\frac{1}{\alpha+2}}\, 
\eta^{\frac{2}{\alpha+2}}, 
\label{BG7}
\end{equation}
which results in the minimum condition 
\begin{equation}
\gamma\ <\ 
f(\rho_{\rm max})\ =\  \frac{\alpha+2}{2\eta^{\frac{2\alpha+2}{\alpha+2}}}\, 
\left( \frac{2}{2\alpha+3} \right) ^{\frac{2\alpha+3}{\alpha+2}}. 
\label{BG8}
\end{equation}
%
 
%==============================================================================
 
\section{Design algorithm} 
\label{Design algorithm} 
At this point we know that the new mechanism is capable of trapping 
dipoles, but several conditions have to be fulfilled simultaneously. 
The following algorithm provides a systematic way for 
designing trap parameters that satisfy the various conditions. 
\begin{enumerate} 
%
% STEP 1
\item We start with the stability criterion (\ref{LSA12}) and use the 
definition of $\beta$ in (\ref{EQMO14}) to obtain the condition 
\begin{equation}
l_0\ <\ \left[ \frac{q_s(\eta)\, \alpha\, I \rho_0^{\alpha} }{M\eta}\right]^{1/2}.
\label{DA1}
\end{equation}
Denote by $D$ the desired location of the radial 
equilibrium distance of the trapped dipole. The size of $D$ 
may depend on many factors, prime among them the spatial 
extent of the field-generating electrodes. If, e.g., the 
field is generated by a wire, we obviously need to choose 
$D>r_{\rm wire}$, where $r_{\rm wire}$
is the radius of the wire. With $D$ chosen, and 
\begin{equation}
D\ =\ \rho_0 l_0, 
\label{DA1a}
\end{equation} 
we now have 
\begin{equation}
\frac{1}{\rho_0^2}\ =\ \frac{l_0^2}{D^2}\ <\ 
\left[ \frac{q_s(\eta)\, \alpha\, I}{M D^2}\right] \, \frac{\rho_0^{\alpha}}{\eta}.  
\label{DA2}
\end{equation}
With (\ref{LSA2a}) this implies 
\begin{equation}
\eta\ >\  \frac{2\, M D^2}{q_s(\eta)\, \alpha(\alpha+1) I}. 
\label{DA3}
\end{equation}
Since the right-hand side of (\ref{DA3}), via $q_s(\eta)$, depends on $\eta$, 
we solve (\ref{DA3}) iteratively. Starting with $q_s(\eta)=0.9$, 
an $\eta$ value satisfying (\ref{DA3}) is usually obtained after 
only a few 
iterations of (\ref{DA3}). 
%
% STEP2
\item With $\eta$ chosen, we may now use (\ref{DA1a}) to compute $l_0$ according to 
\begin{equation}
l_0\ =\ \frac{D}{\rho_0}\ =\ \frac{D} 
{ \left( \frac{\alpha+1}{2}\right)^{\frac{1}{\alpha+2}}\, 
\eta^{\frac{2}{\alpha+2}} }.
\label{DA3a}
\end{equation}
Since $\eta$ satisfies (\ref{DA3}), $l_0$, 
computed according to (\ref{DA3a}), automatically satisfies 
(\ref{DA1}).
%
% STEP 3
\item With $\eta$ chosen and $l_0$ computed, we may now derive a 
condition for the drive frequency $\Omega$. From 
(\ref{BG8}) with (\ref{BG4}) and (\ref{EQMO9}), we obtain 
\begin{equation}
\Omega\ >\  \left[ \frac{8g}{(\alpha+2) l_0}\, 
\left(\frac{2\alpha+3}{2}\right)^{\frac{2\alpha+3}{\alpha+2}} \right]^{1/2}\, 
\eta^{\frac{\alpha+1}{\alpha+2}}. 
\label{DA4}
\end{equation}
%
% STEP 4
\item With $\eta$, $l_0$, and $\Omega$ known, we now use 
(\ref{EQMO10}) to determine $V_{\rm dc}$ according to 
\begin{equation}
V_{\rm dc}\ =\ \frac{M\Omega^2 l_0^{\alpha+2}}{4\alpha p \nu}. 
\label{DA5}
\end{equation}
%

%
% LAST STEP
\item With all design parameters 
chosen, we may now perform an important 
verification step, the 
numerical solution of the system of equations 
\begin{align}
&\ddot \rho\, +\, [1-2\eta\cos(2\tau)]\frac{\cos(\theta)}{\rho^{\alpha+1}}\, -\, \gamma\ =\ 0, 
\label{DAXa}    \\
&\ddot\theta\, +\, \beta [1-2\eta\cos(2\tau)]\frac{\sin(\theta)}{\rho^{\alpha}}\ =\ 0, 
\label{DAXb}
\end{align}
i.e. the system (\ref{EQMO13a}), (\ref{EQMO13b}) including 
the effect of gravity in the radial equation. 
Only if the system (\ref{DAXa}), (\ref{DAXb}) 
exhibits stable solutions did we indeed 
construct a parameter set that leads to the 
stable trapping of point dipoles ${\bm p}$ in the 
field ${\bm E}$. Failure of this verification step 
indicates that the trap parameters were chosen 
too close to the analytical estimates of 
the stability borders. Allowing for larger safety 
margins will cure the problem. 
\end{enumerate} 
%
%==============================================================================

\section{Ferroelectric rods} 
\label{Ferroelectric rods}  
Ferroelectric rods of mass $M$, trapped 
in the electric field of a metallic, spherical 
electrode of radius $r_c$, 
provide an 
illustration of the new trapping mechanism 
that may be realized experimentally. 
To be specific, 
we study the case of thin rods of 
barium titanate, a classic ferroelectric material \cite{Kittel} 
with a spontaneous polarization of 
$P_s=0.15\,$Cm$^{-2}$ \cite{Merz} and 
a density of $6\,$g/cm$^3$ \cite{CRC}. The 
moment of inertia of thin, cylindrical rods is \cite{MTh} 
\begin{equation}
I\ =\ \frac{1}{12}\, M\, b^2,
\label{FER1}
\end{equation}
where $b$ is the length of the rods. We specify 
$b=1\,$mm, $D=2\,$mm, and $\nu=r_c=1\,$mm. 
According to step~1 of the design algorithm and 
$q_s=0.9$, we 
obtain $\eta>17.8$. We choose $\eta=25$, 
which satisfies (\ref{DA3}). 
According 
to step~2 we obtain $l_0=0.36\,$mm. According to 
step~3 we obtain $\Omega>7811\,$s$^{-1}$; we 
choose $\Omega=1.5\times 10^4\,$s$^{-1}$. Step~4, then, 
yields $V_{\rm dc}=18.9\,$V. The checks on 
$\beta$ and $\gamma$ yield 
$\beta=0.78<q_s\rho_0^2/\eta=1.1$ and 
$\gamma=4.8\times 10^{-4}<2(2/7)^{7/4}/\eta^{3/2}=1.8\times 10^{-3}$. 
According to (\ref{LSA2b}) 
the micro-motion amplitude of the rod is 
$\rho_1 l_0=26.6\,\mu$m, which allows enough 
clearance for the rod to oscillate around $r=D$. 
The numerical check of stability according to 
step~5 was performed 
for $\eta=25$, $\beta=0.78$, $\alpha=2$, and $\gamma=4.8\times 10^{-4}$ 
over a time interval of $\Delta\tau=500\pi$ 
for the following $7^4=2401$ initial conditions: $(\rho(\tau=0)=5.3+j\times 0.03, 
\dot\rho(\tau=0)=k\times 0.003, \theta(\tau=0)=l\times 0.01, 
\dot\theta(\tau=0)=m\times 0.01)$, $j,k,l,m=-3,\ldots,3$. 
It was verified graphically that all 2401 trajectories are stable.  
%==============================================================================
 
\section{Discussion} 
\label{Discussion}  
The new ac/dc trapping mechanism proposed here is 
fundamentally different from the saddle-point mechanism 
employed for the successful demonstration of 
electrodynamic trapping of polar molecules 
\cite{VBM,BVSM}. 
The trap proposed here 
makes use of {\it monopole} fields, i.e. particles are 
trapped in close proximity to {\it sources} of 
the electric field. In addition, while the 
pseudo-potential of 
higher-multipole traps generates a {\it focusing} force, the 
pseudo-potential of the trap proposed 
here generates a {\it defocusing} force, 
which attempts to drive 
the particle toward $r=\infty$. Thus, the 
dc voltage 
is an essential ingredient for the 
trap discussed in this paper: the dc voltage 
generates the attractive force counterbalancing 
the defocusing effect of the ac electric field. 
Thus, the trap discussed here is akin to the 
dynamically stabilized inverted pendulum 
\cite{PEND1,PEND2}, 
a classic lecture demonstration in introductory 
mechanics courses. 
  
At a first glance, the new trap introduced here resembles 
a trap investigated several years ago by Riis and Barnett 
\cite{RiisBarnett}. However, there are several important 
differences. 
(a) The Riis/Barnett trap aims to trap molecules 
in the ground state \cite{RiisBarnett}, whereas 
the new trap employs a 
dynamical trapping mechanism in the radial and angular 
degrees of freedom. Quantum mechanically, trapping in 
the angular degree of freedom 
corresponds to a wave function in which 
the trapped particle finds itself in an angularly  
localized wave packet that does not correspond to 
the rotational ground state of the trapped particle. 
(b) For the Riis/Barnett trap gravity is essential 
\cite{RiisBarnett,RBReply}, 
whereas the new trap works even without gravity (see 
Sec.~\ref{NumIn}). 
(c) In contrast to the Riis/Barnett trap, which does 
not use a permanent dipole moment, a permanent 
dipole moment is {\it essential} for  
the new trap to work. 
(d) No stable 
trajectories have been found  \cite{RBReply} for the 
Riis/Barnett trap. 
Far from being trivial, the reason why the Riis/Barnett 
trap does not work is interesting and instructive: 
The micro-motion 
in the Riis/Barnett trap is 
too large \cite{RBReply}. Contrary to the dynamics 
exhibited by the Riis/Barnett trap, 
the micro-motion in the new trap is very small 
(see Fig.~\ref{fig1}), 
explaining 
why the new trap works. In our case, the small micro-motion
(see Fig.~\ref{fig1}) is the reason why 
the pseudo-potential method is valid and leads to 
stable trapping. The importance of a small micro-motion 
for stable trapping and 
for the validity of the pseudo-potential method 
was also pointed out by Cornell \cite{Cornell}. 
   
Our analysis did not include the possibility 
that the trapped dipoles may have orbital angular momentum, i.e., 
the possibility of revolving around the field-generating electrodes. 
Since orbital angular momentum is not an essential part of 
the trapping mechanism, like it is, e.g., in the classic 
Kingdon trap \cite{Kingdon1,orr,Church}, this is not 
an essential omission. On the contrary: It is one of the 
strengths of the new trap that it works for 
zero orbital angular momentum. In fact, 
in the presence of a 
cooling mechanism, zero orbital angular 
momentum will be achieved 
automatically. Zero orbital angular momentum 
is a problem for traps that need 
orbital angular momentum to work (see, e.g., \cite{Seka,SekaSchm}); 
it is not a problem for the new trap. It has been checked that 
the new trap does work in the presence of orbital angular momentum. 
Some aspects of this result are not surprising. 
In the case $\alpha=2$, e.g., 
orbital angular momentum 
creates a centrifugal force with the same radial dependence 
as the dc electric force. Thus 
the effects of angular momentum do not 
structurally change the equations of motion and 
can be counterbalanced 
by an adjustment of ac and dc voltages. 
The fact that the orientation of the dipole 
on its trip around the electrode 
adiabatically follows a near-alignment 
with small oscillations around the orbital radius vector  is 
more of a surprise. This mechanism, however, is not new; it forms 
the dynamical basis of the 
Sekatskii/Schmiedmayer trap \cite{Seka,SekaSchm}, 
which has already been demonstrated to work experimentally 
 \cite{LS}. In fact, the 
Sekatskii/Schmiedmayer trap \cite{Seka,SekaSchm} is 
a limiting case of the new trap 
in the presence of 
orbital angular momentum but zero ac electric field. 
Since the stabilizing repulsive force counteracting the 
attractive dc field is provided by orbital angular momentum 
for zero ac field, and by the repulsive pseudo-force for 
zero orbital angular momentum, 
a ``chirping mechanism'' 
may be envisioned: As the particle loses angular momentum due 
to cooling or damping, an ac voltage may be switched on, providing 
the repulsive force that keeps 
the trapped particle(s) from crashing into the electrode(s). 
The final state is then the zero orbital angular momentum state, 
fully stabilized by the ac field according to the new mechanism. 
   
The trapping mechanism discussed here is also different 
from the mechanism of optical tweezers  \cite{TWEEZE,CHU,BLOCK}. 
Optical tweezers 
work astonishingly well for small particles ranging from 
atoms and molecules to biological cells and even buoyant, small 
organisms \cite{ASH,CHU}. 
However, in contrast to the scheme proposed here, 
optical tweezers may not be able to hold a mm sized 
rod of ferroelectric against gravity, and the frequency of 
optical tweezers ($\sim 10^{15}\,$Hz) is orders of magnitude 
larger than is required according to the 
new trapping scheme. 
Thus, differing in mechanism and range of application, 
each of these two mechanisms has its own optimal 
niche of applicability. 
   
In our analysis above, we used $E\sim 1/r^{\alpha}$ and 
examples were presented for 
$\alpha=2$. Inverse powers is a natural 
choice (wires, e.g.,  generate $\alpha=1$; 
spheres generate $\alpha=2$; oppositely 
charged spheres, e.g., generate $\alpha=3$ for large $r$), 
but not necessary. 
Superpositions of such fields as well as many other 
field configurations may lead to stable trapping, provided 
the corresponding stability criteria are worked out and 
are satisfied. 
 
For the time-dependent part of the electric field 
of the trap we chose a 
$\cos(\Omega t)$ drive (harmonic drive). This, again, 
is not essential. Any drive envelope, 
such as, e.g., rectangular or impulsive (``$\delta$ kicks''), can be accommodated 
and may yield stable, trapped solutions 
following proper adjustment of the $\beta$, $\gamma$, and 
$\eta$ criteria. Impulsive drives, in particular, offer the possibility of 
representing the dynamics of the dipole in the form of 
iterated mappings \cite{Ott}. This 
considerably simplifies the 
analytical and numerical analysis of the trap. 
 
Concerning the case of trapped 
barium titanate rods, we have to be careful 
that the applied electric fields do not 
exceed the coercive field strength, i.e. a
field strength that, applied in the opposite direction, would 
result in a reversal of the direction of the permanent polarization 
of the rods. Since, according to (\ref{EQMO8}), $V_{\rm ac}=2\eta V_{\rm dc}$, 
and $V_{\rm dc}=18.9\,$V in our example, the ac electric field at the 
position $D=2\,$mm is $E_{\rm ac}= r_c V_{\rm ac}/D^2=2400\,$V/cm. 
Since typical coercive fields for barium titanate are on the 
order of $10\,$kV/cm \cite{HULM}, the field 
in our example is tolerable. However, if actual 
experiments are contemplated, it may be advisable to optimize 
the trapping parameters and/or the nature of the trapped particles. 
Rods made of 
croconic acid crystals \cite{CROCO}, e.g., 
may be easier to trap than the barium titanate rods 
discussed in Sec.~\ref{Ferroelectric rods}.
 
Another concern is the large dielectric constant of 
barium titanate, which may result in an appreciable induced 
dipole moment. At room temperature, $\epsilon_r$, the 
relative dielectric constant, may reach values 
of $\epsilon_r\approx 5000$ \cite{Merz}. In this case the induced polarization is 
$P_{\rm ind}=\epsilon_0\epsilon_r E_{\rm ac}=1.1\times 10^{-4}\,$C/m$^2 \ll 
P_s=0.15\,$C/m$^2$, where $\epsilon_0$ is the permittivity of the vacuum. 
We conclude that induced polarization is not 
an important effect. Even if for some trapped particle 
species this effect should be large, the induced polarization 
may be fully compensated for by adjusting 
$V_{\rm ac}$ and $V_{\rm dc}$.   
 
While this paper focuses on the case of a single stored 
dipole, we may also study the case of many, simultaneously 
stored dipoles. In the presence of a cooling mechanism, such as 
laser cooling \cite{Sten} (for atomic-sized particles) or 
buffer-gas cooling \cite{Buffer} (for atomic-, nano-, and macroscopic-sized 
particles), trapped particle densities may become large enough that 
dipole-dipole interactions 
are important. In the case of molecules this may allow us to 
study new 
ferroelectric phases of dilute neutral (trapped) molecular gases. 
In the case of 
small ferroelectric particles, in analogy to the geometric arrangements 
of trapped Coulomb crystals \cite{MPQ-PRL,NIST-PRL,RB-Nature},  
a geometrically ordered dipolar phase may 
be generated and studied. This may result in an interesting contribution to the emerging field of 
granular materials \cite{GranMat}. 
%==============================================================================

\section{Summary and conclusions} 
\label{Summary and conclusions}  
  
This paper presents a
novel electrodynamic trap for the stable confinement of 
neutral particles with 
permanent electric dipole moments. The mechanism of the trap differs 
profoundly from currently demonstrated or proposed neutral-particle 
traps. While conventional electric traps are based on energy shifts 
in selected quantum states 
and employ higher-multipole fields as an essential 
ingredient of trapping, the new trap uses a 
superposition of {\it monopole} fields, i.e. 
an attractive, static dc electric field and 
an ac field, whose net effect results in strong repulsion. 
If certain conditions on the strengths of the dc and ac fields are met, 
stable trapping results at the point where attractive and repulsive 
forces are in equilibrium. A mechanical analogue of this 
trapping mechanism is the dynamically stabilized inverted pendulum. 
Stability of the trap is proved with the help of an analytical 
stability analysis, supplemented with detailed numerical 
molecular trajectory simulations. 
The effect of gravity is considered and included. The 
example of 
thin barium titanate rods indicates that the trap works in practice. 
It is possible that the trap also works for polar (macro) molecules. 
Although there are no magnetic charges, and magnetic 
fields are divergence-free, it is possible that 
close to a magnetic pole as a field generating 
``source'' a superposition 
of static and oscillating magnetic fields may result in a magnetic 
analog of the trapping mechanism discussed here. 
%==============================================================================


\begin{thebibliography}{99}
 
\bibitem{HRSpec1} H. G. Dehmelt, in {\it Advances in Laser Spectroscopy} 
              (edited by F. T. Arecchi, F. Strumia, and H. Walther), 153--187 
              (Plenum, NY, 1983). 
 
\bibitem{HRSpec2} D. J. Wineland, W. M. Itano, and R. S. Van Dyck, in 
               {\it Adv. Atom. Molec. Phys.}, Vol. {\bf 19} 
               (edited by D. R. Bates and B. Bederson), 135--186 
               (Academic, NY, 1983). 
 
\bibitem{HRSpec3} D. J. Wineland and W. M. Itano, Phys. Lett. {\bf 82A}, 
               75 
%          -- 78 
              (1981). 
 
\bibitem{FrequStand} H. G. Dehmelt, IEEE Trans. Instrumn. Meast. 
               {\bf IM-31}, 83 
%     -- 87 
              (1982). 
 
\bibitem{NAM1} M. Amoretti {\it et al.}, 
% Amsler C, Bonomi G, Bouchta A, Bowe P, Carraro C, Cesar CL, Charlton M, Collier MJ, Doser M, Filippini V, 
% Fine KS, Fontana A, Fujiwara MC, Funakoshi R, Genova P, Hangst JS, Hayano RS, Holzscheiter MH, 
% J¿rgensen LV, Lagomarsino V, Landua R, Lindelšf D, Lodi Rizzini E, Macr" M, Madsen N, Manuzio G, 
% Marchesotti M, Montagna P, Pruys H, Regenfus C, Riedler P, Rochet J, Rotondi A, Rouleau G, Testera G, 
% Variola A, Watson TL, van der Werf DP, 
%
%   {\it Production and detection of cold antihydrogen atoms}, 
              Nature {\bf 419}, 456 
%        -- 459 
      (2002).   
 
\bibitem{NAM2} J. N. Tan {\it et al.}, 
%   {\it Observations of cold antihydrogen}, 
          Nucl. Instr. Meth. Phys. Res. B {\bf 214}, 
          22 
%     -- 30 
       (2004). 
 
\bibitem{JILABEC} M. H. Anderson, J. R. Ensher, M. R. Matthews, C. E. Wieman, and E.A. Cornell,
%         {\it Observation of Bose-Einstein Condensation in a Dilute Atomic Vapor}, 
                Science {\bf 269}, 198
%             --201
              (1995).   %@

\bibitem{MITBEC} K.B. Davis, M.O. Mewes, M.R. Andrews, N.J. van Druten, D.S. Durfee, D.M. Kurn, 
                                 and W. Ketterle,  
%            {\it Bose-Einstein Condensation in a Gas of Sodium Atoms},            
               Phys. Rev. Lett. {\bf 75}, 3969
%             Ð3973          
             (1995).
 
\bibitem{QComp} {\it Quantum Information Processing}, 
        edited by G. Leuchs and Th. Beth (Wiley-VCH, Weinheim, 2003). %@
    
\bibitem{MagLev} M. Tsuchiya and H. Ohsaki, 
% {\it Characteristics of electromagnetic force of EMS-type maglev 
%       vehicle using bulk superconductors}, 
               IEEE Transactions on Magnetics {\bf 36}, 
               3683
%                -- 3685 
               (2000).   %@
 
\bibitem{Kingdon1} K. H. Kingdon,
                  Phys. Rev. {\bf 21}, 408 (1923).%%

\bibitem{Penning} F. M. Penning, Physica {\bf 3}, 873 (1936). 
              
\bibitem{Paul1}  W. Paul, O. Osberghaus, and E. Fischer,
                 Forschungsber.
                 Wirtsch. Verkehrsminist.
                 Nord\-rhein-Westfalen {\bf 415},
                 1 (1958). 
                 
\bibitem{orr} T. Biewer, D. Alexander, S. Robertson,
              and B. Walch, Am. J. Phys.
              {\bf 62}, 821 (1994).  
    
\bibitem{Church} D. A. Church, 
%  {\it Collision measurements and excited-level lifetime measurements on ions stored in 
%           Paul, Penning and Kingdon ion traps}, 
Phys. Rep. {\bf 228},
                 253
% - 258
         (1993). 

\bibitem{MPQ-PRL} F. Diedrich, E. Peik, J. M. Chen, W. Quint, and 
                                    H. Walther, 
%    {\it Observation of a Phase Transition of Stored Laser-Cooled Ions}, 
Phys. Rev. Lett. {\bf 59}, 2931
%  - 2934 
(1987). %@
 
\bibitem{NIST-PRL} D. J. Wineland, J. C. Bergquist, W. M. Itano, 
                                    J. J. Bollinger, and C. H. Manney, 
%  {\it Atomic-Ion Coulomb Clusters in an Ion Trap}, 
Phys. Rev. Lett. {\bf 59}, 2935 
%  - 2938 
  (1987).   %@
 
\bibitem{RB-Nature} R. Bl\"umel, J. M. Chen, E. Peik, W. Quint, 
               W. Schleich, Y. R. Shen, and H. Walther, 
%  {\it Phase transitions of laser-cooled ions}, 
Nature {\bf 334}, 309 
%  - 313 
  (1988). %@
%
\bibitem{WHW} W. H. Wing, 
% {\it Electrostatic Trapping of Neutral Atomic Particles}, 
Phys. Rev. Lett. {\bf 45}, 631 
%  -- 634
  (1980).   %@
%
\bibitem{Seka} S. K. Sekatskii, 
%  {\it Electrostatic traps for polar molecules}, 
JETP Lett. {\bf 62}, 916
%  -- 920 
  (1995).  %@
 
\bibitem{SekaSchm} S. K. Sekatskii and J. Schmiedmayer, 
%  {\it Trapping polar molecules with a charged wire}, 
   Europhys. Lett. {\bf 36}, 407
%   --412 
     (1996).   %@
      
%
\bibitem{BethNat} 
H. L. Bethlem, G. Berden, F. M. H. Crompvoets, 
R. T. Jongma, A. J. A. van Roij, and G. Meijer, 
%  {\it Electrostatic trapping of ammonia molecules}, 
Nature {\bf 406}, 491 
%  -- 494 
 (2000).  %@
%
\bibitem{VBM} 
J. van Veldhoven, H. L. Bethlem, and G. Meijer, 
% \it{ac Electric Trap for Ground-State Molecules}, 
Phys. Rev. Lett. {\bf  94}, 083001 (2005).  %@
%
\bibitem{BVSM} 
H. L. Bethlem, J. van Veldhoven, M. Schnell, and G. Meijer, 
%  {\it Trapping polar molecules in an ac trap}, 
Phys. Rev. A {\bf 74}, 063403 (2006).  %@
 
\bibitem{MWtrap} D. DeMillea , D.R. Glenn, and J. Petricka, 
%  {\it Microwave traps for cold polar molecules}, 
Eur. Phys. J. D {\bf 31}, 375
% Ð384 
  (2004). %@
   
\bibitem{Lovelace} R. V. E. Lovelace, C. Mehanian, T. J. Tommila, and D. M. 
Lee, 
% {\it Magnetic confinement of a neutral gas}, 
Nature {\bf 318}, 30 
% - 36
 (1985).  %@
 
\bibitem{Berge} A. L. Migdall, J. V. Prodan, W. D. Phillips, T. H. Bergeman, and 
H. J. Metcalf, 
%  {\it First Observation of Magnetically Trapped Neutral Atoms}, 
Phys. Rev. Lett. {\bf 54}, 2596
% - 2599 
  (1985). %@
 
\bibitem{BIO} K. C. Neuman, E. H. Chadd, G. F. Liou, K. Bergman, and S. M. Block,  
%           {\it Characterization of photodamage to escherichia coli in optical traps}, 
              Biophys. J. {\bf 77}, 2856
%          -- 2863 
                (1999).
 
\bibitem{ASH} A. Ashkin and J. M. Dziedzic, 
%     {\it Optical trapping and manipulation of 
%          viruses and bacteria},
          Science {\bf 235}, 1517
%      -- 1520 
         (1987). 
  
\bibitem{LL}  L. D. Landau and E. M. Lifshitz, {\it Mechanics}
                    (Pergamon, Oxford, 1960).  %@
         
\bibitem{Ott} E. Ott, {\it Chaos in Dynamical Systems} 
             (Cambridge University Press, Cambridge, 1993).    %@
  
\bibitem{AS}   M. Abramowitz and I. A. Stegun,
               {\it Handbook of Mathematical Functions}
               (National Bureau of Standards, Washington DC,
                1964).  %@
  
\bibitem{Dehmelt}  H. G. Dehmelt,
                   Adv. Atom. Mol. Phys. {\bf 3},
                   53 (1967).

\bibitem{Friedrich} H. Friedrich, {\it Theoretical Atomic Physics}, 
               third edition (Springer, Berlin, 2006).   %@
  
\bibitem{Morales} B. M. Lamb and G. J. Morales, 
%           {\it Ponderomotive effects in nonneutral plasmas}, 
               Phys. Fluids {\bf 26}, 3488 (1983). 
  
\bibitem{K1} R. Bl\"umel, 
%          {\it Dynamic Kingdon Trap}, 
             Phys. Rev. A
             {\bf 51}, R30
%          -- R33
               (1995).  %@
 
\bibitem{K6} I. Garrick-Bethell, Th. Clausen, and R. Bl\"umel,
%  {\it Universal instabilities of radio-frequency traps},
             Phys. Rev. E {\bf 69}, 056222 (2004).  %@
 
%\bibitem{MWSpec} A. Honig, M. Mandel, M. L. Stitch, and C. H. Townes, 
%          {\it Microwave Spectra of the Alkali Halides} 
%              Phys. Rev. {\bf 96}, 629 
%          -- 642 
%             (1954).   %@ 
                   
\bibitem{Kittel} C. Kittel, {Introduction to Solid State Physics} 
              (John Wiley \& Sons, New York, 1971), p. 476.   %@
                   
\bibitem{Merz} W. J. Merz, 
%           {\it The Electric and Optical Behavior of BaTiO$_3$ 
%                 Single-Domain Crystals}
                Phys. Rev. {\bf 76}, 1221
%            -- 1225
                (1949).              %@
   
\bibitem{CRC} {\it Handbook of Chemistry and Physics}, 55th Edition, 
              1974--1975, edited by R. C. Weast (CRC Press, 1974), 
              p. B-72.   %@
   
\bibitem{MTh} J. B. Marion and S. T. Thornton, 
                          {\it Classical Dynamics}, third edition 
                          (Harcourt Brace Jovanovich, Fort Worth, 1988). %@
  
%\bibitem{Paul2} W. Paul,
%                Rev. Mod. Phys. {\bf 62}, 531 (1990). 

\bibitem{PEND1} J. G. Fenn, D. A. Bayne, and B. D. Sinclair, 
%       {\it Experimental investigation of the ``effective potential'' of 
%             an inverted pendulum},
          Am. J. Phys. {\bf 66}, 981 (1998). 

\bibitem{PEND2} E. I. Butikov, 
%       {\it On the dynamic stabilization of an inverted pendulum},
          Am. J. Phys. {\bf 69}, 755 (2001). 
 
\bibitem{RiisBarnett} E. Riis and S. M. Barnett, 
%   {\it A Dynamic Electric Trap for Ground-State Atoms}, 
   Europhys. Lett. {\bf 21}, 533 
%  -- 538 
    (1993).  %@
   
\bibitem{RBReply} E. Riis and S. M. Barnett, 
%   {\it Reply to the Comment of E. A. Cornell}, 
   Europhys. Lett. {\bf 30}, 441
% Reply consists of exactly one page (page 441) 
   (1995).  %@
 
\bibitem{Cornell}  E. A. Cornell, 
%   {\it Comment on ``A Dynamic Electric Trap for Ground-State Atoms''},
    Europhys. Lett. {\bf 30}, 439 
%  -- 440 
  (1995).  %@
 
\bibitem{LS} H. J. Loesch and B. Scheel, 
%  {\it Molecules on Kepler Orbits: An Experimental Study} 
   Phys. Rev. Lett. {\bf 85}, 2709 
%  - 2712 
   (2000).   %@
 
\bibitem{TWEEZE}  A. Ashkin, J.M. Dziedzic, J.E. Bjorkholm, and S. Chu, 
%        {\it Observation of a Single-Beam Gradient Force Optical Trap for Dielectric Particles}, 
           Opt. Lett. {\bf 11} 288
%          --290 
           (1986).     
 
\bibitem{CHU} S. Chu, 
%   {\it Laser Trapping of Neutral particles}, 
       Sci. Am. {\bf 266}, 70 
%   -- 76 
       (February, 1992).   %@. 
 
\bibitem{BLOCK}  S. M. Block, 
%    {\it Making light work with optical tweezers}, 
       Nature {\bf 360}, 493 
%   -- 495 
      (1992).   
 
\bibitem{HULM} J. K. Hulm, 
%     {\it Dielectric Properties of Single Crystals of Barium Titanate}, 
        Nature {\bf 160}, 127 
%    -- 128 
        (1947).    %@
   
\bibitem{CROCO} S. Horiuchi, Y. Tokunaga, G. Giovannetti, S. Picozzi, H. Itoh, R. Shimano, 
                                 R. Kumai, and Y. Tokura, 
% {\it Above-room-temperature ferroelectricity in a 
%       single-component molecular crystal}, 
                Nature {\bf  463}, 789
%  -- 792 
              (2010). 

\bibitem{Sten} S. Stenholm,
               Rev. Mod. Phys {\bf 58}, 699 (1986).

\bibitem{Buffer} R. V. Krems, D. Zgid, G. Chalasinski, J. Klos, and A. Dalgarno, 
% {\it Possibility of buffer-gas cooling of paramagnetic carbon to ultracold temperatures}, 
            Phys. Rev. A {\bf 66}, 030702(R) (2002).   %@

\bibitem{GranMat} A. Mehta, {\it Granular Matter: An Interdisciplinary Approach} 
              (Springer, New York, 1994).    %@

\end{thebibliography}
\end{document}